\documentclass[aps,twocolumn]{revtex4}%
\usepackage{amsfonts}
\usepackage{amsmath}
\usepackage{amssymb}
\usepackage{graphicx}%
\providecommand{\U}[1]{\protect\rule{.1in}{.1in}}

\begin{document}
\title{Interplay between s-d exchange interaction and Rashba effect: spin-polarized transport}
\author{W. Yang }
\author{Kai Chang}
\altaffiliation[Author to whom the correspondence should be addressed. Electronic address: ]{kchang@red.semi.ac.cn}

\author{X.G. Wu, and H.Z. Zheng}
\affiliation{SKLSM, Institute of Semiconductors, Chinese Academy of Sciences, P. O. Box
912, 100083, Beijing, China}
\author{F. M. Peeters}
\affiliation{Department of Physics, University of Antwerp (Campus Drie Eiken), B-2610
Antwerp, Belgium}

\begin{abstract}
We investigate the spin-polarized transport properties of a two-dimensional
electron gas in a n-type diluted magnetic narrow gap semiconductor quantum
well subjected to a perpendicular magnetic and electric field. Interesting
beating patterns in the magneto resistance are found which can be tuned
significantly by varying the electric field. A resonant enhancement of
spin-polarized current is found which is induced by the competition between
the \textit{s-d} exchange interaction and the Rashba effect [Y. A. Bychkov and
E. I. Rashba, J. Phys. C \textbf{17}, 6039 (1984)].

\end{abstract}

\maketitle

Recently, spin related phenomena in semiconductors have acquired a renewed
interest because of the possibility of using the spin degree of freedom of
carriers to construct spintronic devices.\cite{Prinz} One of the central
issues in semiconductor spintronics is the creation of spin polarized current.
In a diluted magnetic semiconductor (DMS), the large \textit{s-d} exchange
interaction between the conduction electrons and the localized magnetic ions
provides the possibility of tailoring the electron spin splitting and
consequently modify significantly the transport property of spin polarized
electrons. Via vertical tunnelling through DMS and ferromagnetic metallic
junctions, efficient spin injection into semiconductors has been
demonstrated.\cite{Wolf,Oestreich,Fiederling,Ohno,FM1,FM2} The motion of an
electron spin in a DMS or conventional semiconductor may also be modified by
the so-called Rashba spin-orbit interaction (SOI).\cite{Rashba} This
interaction offers the opportunity to manipulate electron spin in a
two-dimensional electron gas (2DEG) via an external electric field instead of
a magnetic field.\cite{Datta} The coupling may lead to interesting effects,
e.g., the recently demonstrated\cite{SHE,SHE2} spin-Hall
effect.\cite{SpinHall} Experimentally, it was found that the Rashba SOI can
modify significantly the magneto-transport property of 2DEG in a InGaAs/InAlAs
heterostructure.\cite{Nitta}

In this Letter, we investigate theoretically the spin transport properties of
a 2DEG in a paramagnetic DMS in the presence of a perpendicular magnetic and
electric field. We include both the \textit{s-d} exchange interaction, which
can be tuned by the magnetic field, and the Rashba SOI, whose strength can be
modified by a perpendicular electric field. We found that a strong
resonant-enhanced spin-polarized current may appear at a critical Rashba SOI
strength even at low magnetic fields.

Let us consider a 2DEG confined in a Mn-based DMS quantum well in the presence
of perpendicular magnetic and electric fields. The Hamiltonian of the
conduction electron including the SOI and the \textit{s-d} exchange
interaction\cite{Furdyna} is given by%

\begin{equation}
\hat{H}=\frac{(\mathbf{p}+e\mathbf{A})^{2}}{2m_{e}}+\frac{\alpha}{\hbar
}[(\mathbf{p}+e\mathbf{A})\times\mathbf{e}_{z}]\cdot\mathbf{\sigma}+\hat
{H}_{z}+\hat{H}_{ex}, \label{hamiltonian}%
\end{equation}
where $\alpha$ is the Rashba SOI strength, $\hat{H}_{z}=g_{e}\mu_{B}%
B\hat{\sigma}_{z}/2$ is the intrinsic Zeeman term, and $\hat{H}_{ex}%
=-\sum\nolimits_{j}J(\mathbf{r}-\mathbf{R}_{j})\mathbf{\sigma}/2\cdot
\mathbf{S}_{j}$ is the \textit{s-d} exchange interaction between the
conduction electron spin $(\mathbf{\sigma}/2)$ and the localized spins
$(\mathbf{S}_{j})$. Within the mean-field approximation, $\hat{H}_{ex}%
=-N_{0}\alpha_{ex}x\left\langle S_{z}\right\rangle \hat{\sigma}_{z}/2,$ where
$N_{0}\alpha_{ex}$ describes the strength of the \textit{s-d} exchange
interaction, $x$ is the fractional occupancy of the Mn ions on cation sites,
and $\left\langle S_{z}\right\rangle $ is the thermal average of the Mn spin.

The eigenvalues and eigenstates of the electron are $(n=1,2,\cdots)$%
\begin{align}
E_{n}^{(\pm)}  &  =(n\pm\delta)\hbar\omega_{c},\text{ \ }E_{0}=(\hbar
\omega_{c}-\Delta_{Z})/2,\label{eigenvalue}\\
\delta &  =\sqrt{[\Delta_{Z}/(\hbar\omega_{c})-1]^{2}/4+2n[\alpha/(l_{B}%
\hbar\omega_{c})]^{2}},\nonumber
\end{align}%
\begin{align}
\psi_{n,k_{y}}^{(\pm)}  &  =e^{ik_{y}y}/\sqrt{L_{y}}[C_{n}^{(\pm)}\phi
_{n-1}\left\vert \uparrow\right\rangle +D_{n}^{(\pm)}\phi_{n}\left\vert
\downarrow\right\rangle ],\label{eigenstate}\\
\psi_{0,k_{y}}  &  =e^{ik_{y}y}/\sqrt{L_{y}}\phi_{0}\left\vert \downarrow
\right\rangle . \label{n0downstate}%
\end{align}
Here $\Delta_{Z}=g_{e}\mu_{B}B-N_{0}\alpha_{ex}x\left\langle S_{z}%
\right\rangle $, $L_{y}$ is the sample length along the $y$ axis, $\{\phi
_{n}\}$ are harmonic oscillator functions centered at $x_{0}=-k_{y}l_{B}^{2}$
[$l_{B}=\sqrt{\hbar/(eB)}$], $C_{n}^{(+)}=-D_{n}^{(-)}=P_{n}/A_{n},$
$C_{n}^{(-)}=D_{n}^{(+)}=1/A_{n}$ for $\Delta_{Z}<\hbar\omega_{c}$,
$C_{n}^{(+)}=D_{n}^{(-)}=1/A_{n}$, $C_{n}^{(-)}=-D_{n}^{(+)}=-P_{n}/A_{n}$ for
$\Delta_{Z}>\hbar\omega_{c}$, $P_{n}=\sqrt{2n}l_{B}^{-1}\alpha/(\left\vert
\Delta_{Z}-\hbar\omega_{c}\right\vert /2+\hbar\omega_{c}\delta)$, and
$A_{n}=\sqrt{1+P_{n}^{2}}$. It is interesting to notice that the zeroth Landau
level (LL)\ ($\psi_{0,k_{y}}$) is decoupled from other LL's, and is always a
spin-down eigenstate when the strength of the Rashba SOI ($\alpha$) varies.

The broadening of the LL's induced by impurity scattering is obtained from the
Fermi golden rule $\Gamma_{nk_{y}\lambda}=\hbar\sum_{n^{\prime}k_{y}^{\prime
}\lambda^{\prime}}W_{nk_{y}\lambda,n^{\prime}k_{y}^{\prime}\lambda^{\prime}%
}\ (\lambda,\lambda^{\prime}=\pm)$, where $W_{i,j}$ is the impurity scattering
rate from initial state $\left\vert i\right\rangle $ to final state
$\left\vert j\right\rangle $. Based on the quantum Boltzmann
equation,\cite{Vasilopoulos} the conductivity is given by%

\begin{equation}
\sigma_{xx}=\frac{e^{2}}{2k_{B}T\Re}\sum\limits_{i,j}f(E_{i})\left[
1-f(E_{j})\right]  W_{ij}(\left\langle x\right\rangle _{i}-\left\langle
x\right\rangle _{j})^{2}, \label{ConductivityXX}%
\end{equation}
where $f(E_{i})$ is the Fermi distribution function, $\left\langle
x\right\rangle _{i}$ is the cyclotron center of state $\left\vert
i\right\rangle $. The resistivity is obtained from $\rho_{xx}=\sigma_{xx}/S,$
where $S=(\sigma_{xx})^{2}+(\sigma_{xy})^{2}\approx(en_{e}/B)^{2}$.

In the presence of the SOI, the eigenstates of the electron are no longer pure
spin-up or spin-down states, but a mixture of them (except $\psi_{0,k_{y}}$).
The most widely used definition of the current polarization, is $P_{J}%
=(\sigma_{xx}^{\uparrow}-\sigma_{xx}^{\downarrow})/(\sigma_{xx}^{\uparrow
}+\sigma_{xx}^{\downarrow}),$\cite{Loss} with%

\begin{equation}
\sigma_{xx}^{\uparrow,\downarrow}=\eta\sum\limits_{i,j}f(E_{i})\left[
1-f(E_{j})\right]  W_{ij}(\left\langle x\right\rangle _{i}-\left\langle
x\right\rangle _{j})^{2}P_{j}^{\uparrow,\downarrow}, \label{sigma_Spin}%
\end{equation}
where $\eta=e^{2}/(2k_{B}T\Re)$, $P_{j}^{\uparrow}=|C_{j}^{\pm}|^{2}$
($P_{j}^{\downarrow}=|D_{j}^{\pm}|^{2}$) is the spin-up (spin-down)
probability of the final state $\left\vert j\right\rangle $. The definition of
the current polarization $P_{J}$ is consistent with the usual definition in
the case of zero SOI ($\alpha=0$), and approaches zero in the case of strong
SOI ($\alpha\rightarrow\infty$) as expected. In the case of $\alpha=0$, our
theoretical model can produce excellent agreement with recent experiment
demonstrating the validity of our theory.\cite{Teran}.

\begin{figure}[ptb]
\includegraphics[width=\columnwidth]{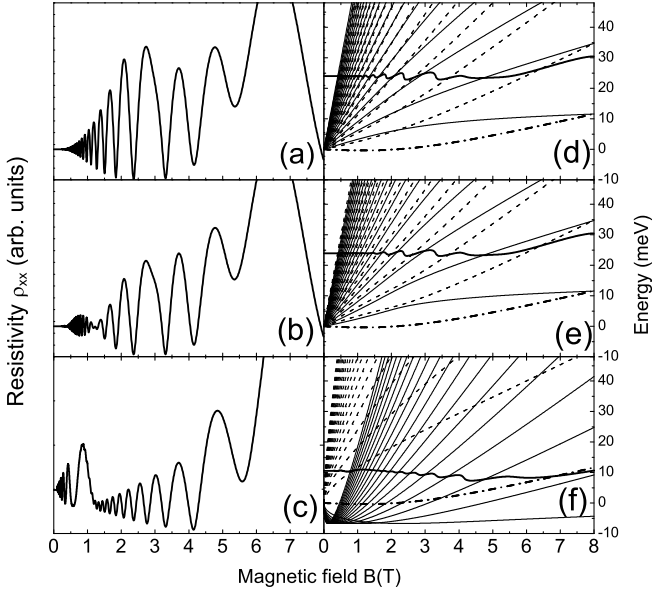}\caption{Longitudinal
magnetoresistivity $\rho_{xx}$ (a,b,c) and energy spectra (d,e,f) at $T=1$ K
as a function of magnetic field for $\alpha$= 0, 5, 160 meV nm. The thickest
solid lines in Figs. 1(d,e,f) denote the Fermi energies, the thin solid and
dashed lines denote the (-) and (+) branches of LL's, respectively. The
thicker dash-dotted lines denote the zeroth spin-down LL.}%
\end{figure}In order to investigate the interference between the \textit{s-d}
exchange interaction and the Rashba effect, we consider a narrow bandgap DMS
2DEG in Hg$_{0.88}$Mn$_{0.12}$Te, where the Rashba SOI strength could be
strong.\cite{alpha} The following parameters are used: $E_{g}=0.2$ eV,
$N_{0}\alpha_{ex}=400\ $meV,\ $g_{e}=-41$,\ $x_{eff}=0.02$,\ $m_{e}=0.04m_{0}$
($m_{0}$ is the free electron mass), $n_{I}=3\times10^{9}\ $cm$^{-2}$, and
electron density $n_{e}=4\times10^{11}\ $cm$^{-2}$ \cite{Furdyna,Gui}. The
magnetoresistance $\rho_{xx}$ is shown in Figs. 1 (a,b,c) for different
perpendicular electric field, i.e., different Rashba SOI strengths
($\alpha=0,5,160\ $meV nm). The beating patterns, which are imposed on the
Shubnikov-de Haas oscillations of $\rho_{xx}$, arise from the sweep of the (-)
branch LL's over the (+) branch LL's. The behavior of $\rho_{xx}$ is changed
significantly by introducing the Rashba SOI [see Fig. 1(e,f)], since the
interplay between the s-d exchange interaction and the Rashba SOI\cite{CK}
changes the relative position of the two branches of LL's [see Eq.
(\ref{eigenvalue})]. For very strong Rashba SOI, $\rho_{xx}$ exhibits a very
different behavior due to the separation of the two branches ($\pm$) of LL's
[see Figs. 1(c,f)].

\begin{figure}[ptb]
\includegraphics[width=\columnwidth]{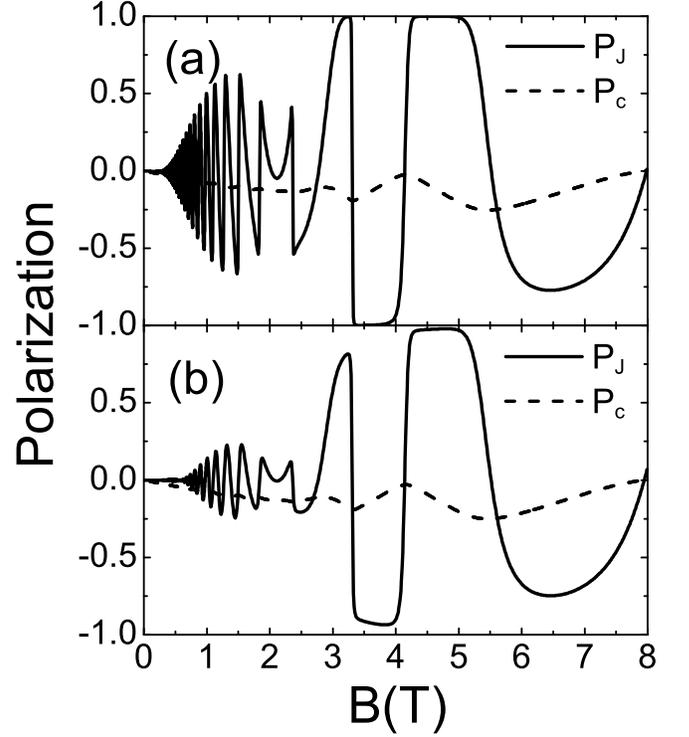}\caption{Spin polarization of
carrier $P_{c}$ and current $P_{J}$ for (a) $\alpha=0$ and (b) $\alpha=5$ meV
nm at T=1 K.}%
\end{figure}In the absence of the Rashba SOI, a strong polarization of the
current can be obtained by applying a strong magnetic field [see Fig. 2(a)].
It is interesting to find that the current polarization $P_{J}$ oscillates
with increasing magnetic fields and approaches $\pm$100\% at strong magnetic
fields while the carrier polarization $P_{c}\equiv(n_{\uparrow}-n_{\downarrow
})/(n_{\uparrow}+n_{\downarrow})$ is always much smaller. The physical reason
is that $P_{J}$ is related to the density-of-states polarization on the Fermi
surface, while $P_{c}$ is determined by the carrier population. Thus our
theoretical results show that a spin-polarized current in the DMS 2DEG can be
realized by adjusting the magnetic field or electron density such that $E_{F}$
lies at the center of one spin-split LL.

\begin{figure}[ptb]
\includegraphics[width=\columnwidth]{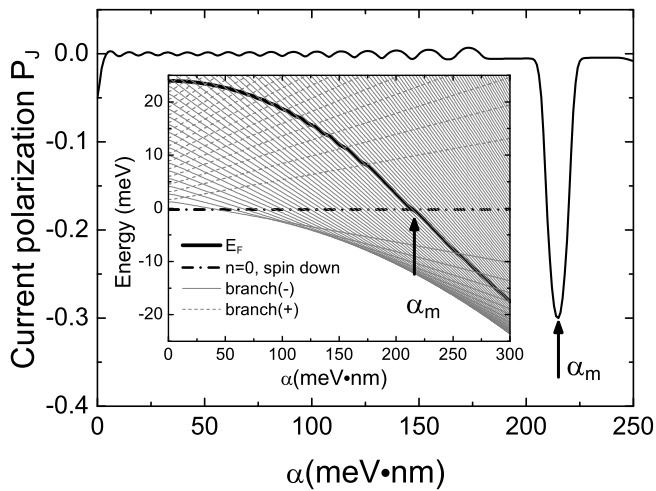}\caption{Spin polarization of
current vs. the SOI strength for a fixed B=0.5 T at T=1 K. The inset shows the
LL's and the Fermi energy as a function of the SOI strength. The line type is
the same as Fig. 1. $n_{e}=4\times10^{11}$ cm$^{-2}$, and $x_{eff}=0.02$.}%
\end{figure}The influence of Rashba SOI behaves like a momentum-dependent
in-plane effective magnetic field [see Eq. (\ref{hamiltonian})]. It mixes the
spin-up and the spin-down LL's except ($\psi_{0,k_{y}}$) such that the
electron eigenstates are no longer spin eigenstates, which, in most
situations, reduces the spin polarization of the current [see Fig. 2(b)].
However, it is interesting to notice that the Rashba SOI could induce a
resonant enhancement of the spin polarization of the current (see the deep dip
at $\alpha_{m}$ in Fig. 3) for a fixed magnetic field. This interesting
feature can be understood from the inset of Fig. 3. For strong Rashba SOI, the
two branches ($\pm$) of LL's consist of nearly equal spin-up and spin-down
components, leading to a vanishingly small current polarization [see Eq.
(\ref{sigma_Spin})] when they coincide with the Fermi energy. In contrast, the
zeroth spin-down LL is a spin-down eigenstate and it contributes a pure
spin-down current when it coincides with $E_{F}$. With increasing Rashba SOI,
branch (-) [(+)] LL's are shifted to lower (higher) energies while the energy
of the zeroth spin-down LL keeps constant [see Eq. (\ref{eigenvalue})].
Consequently, the Fermi energy decreases and coincides with the zeroth
spin-down LL, leading to resonant enhancement of the spin polarization of the current.

In conclusion, the competition between the \textit{s-d} exchange interaction
and the Rashba SOI provides us with an interesting possibility to create
spin-polarized current in low-dimensional semiconductor structures. Strong
spin-polarized current can be achieved even when the polarization of carriers
is very small. A large spin polarization of current induced by the Rashba
effect is predicted at low magnetic field.

\acknowledgments This work is partly supported by the NSF and MOST of China,
and the Bilateral Cooperation programme between Flanders and China.

\end{document}